\documentclass[]{spie}  

\usepackage{amsmath,amsfonts,amssymb}
\usepackage{graphicx}
\usepackage{lineno}
\usepackage[colorlinks=true, allcolors=blue]{hyperref}

\title{Architecture and validation of the CRS F-Engine for the CHORD radio telescope}

\author[a,b,c]{Ian Hendricksen}
\author[a,c]{Jean-François Cliche}
\author[a,b,c]{Matt Dobbs}
\author[c]{Joshua Montgomery}
\affil[a]{McGill University Department of Physics, 3600 rue University, Montréal, Canada}
\affil[b]{Trottier Space Institute, 3550 rue University, Montréal, Canada}
\affil[c]{t0.technology, 300 rue Léo-Pariseau, Montréal, Canada}

\authorinfo{Further author information: (Send correspondence to I.H.)\\I.H.: E-mail: ian.hendricksen@mail.mcgill.ca}

\begin{document} 
\maketitle

\begin{abstract}

We introduce the design of the t0.technology Control and Readout System (CRS) F-Engine that will be used for the Canadian Hydrogen Observatory and Radio transient Detector (CHORD), a new radio interferometer currently being commissioned at the Dominion Radio Astrophysical Observatory (DRAO) in Canada. The CRS F-Engine will directly digitize and channelize $1024$~individual RF signals from the $512$~dual-polarized dishes of the core array using an array of 128 CRS boards, a multi-purpose microwave readout platform using an AMD Zynq Ultrascale+ RF-System-on-Chip (RFSoC) architecture. The CRS supports the required analog and digital signal processing and is appropriately scalable, with rack-mountable crates each supporting up to $16$~CRS boards, equipped with a backplane for distribution of power, common clock and time synchronization signals, and a full-mesh network for intra-crate data transmission. Implemented on the CRS boards is the \texttt{chFPGA} firmware which supports the digitization of $8$~analog signals at $3.2$~GSPS and channelizes them with a CASPER-based PFB/FFT into $8,192$~frequency bins with $\sim 195$~kHz of resolution, which are then re-quantized into $(4 + 4i)$~bits for data offload to an external X-Engine. \texttt{chFPGA} supports multiple post-channelization signal processing options through separate bitstream files for different applications, such as a $100$~GbE packet assembler-transmitter for CHORD to feed channelized data to its external GPU-based X-Engine, as well as FPGA-based $N^2$ correlators, including a single-board $(N = 8)$ correlator (the ``Pocket Correlator"), and a multi-board corner-turn engine coupled with a half-CRS crate $(N = 64)$ correlator. We demonstrate the performance of \texttt{chFPGA} by injecting a wideband Gaussian noise source into a CRS board running the Pocket Correlator firmware, and find that recovered digitized timestream and channelized data are in excellent agreement with expectations.

\end{abstract}


\section{INTRODUCTION}\label{sec:intro}

We present the design of the t0.technology Control and Readout System (CRS) F-Engine for synchronous digitization and efficient channelization of radio signals over a $1.6$~GHz bandwidth, featuring a fully-scalable architecture supporting $\mathcal{O}$(1--1000) inputs. We introduce the CRS F-Engine as the selected F-Engine platform for the Canadian Hydrogen Observatory and Radio transient Detector (CHORD), a next-generation large-$N$, small-$D$ radio interferometer expected to achieve significant advances in $21$-cm intensity mapping, discovery of fast radio bursts (FRBs), pulsar surveys, and spectral line galaxy surveys throughout the next decade \cite{CHORDWhitepaper, Dobbs2026CHORDOverview}. In this paper, we describe the hardware, firmware and software architecture of the CRS F-Engine, and validate its performance by using a stand-alone single-board F-X correlator firmware configuration of the same platform. In Section~\ref{sec:intro}, we provide an overview of CHORD and introduce the CRS F-Engine hardware. In Section~\ref{sec:firmware}, we describe the digital signal processing firmware \texttt{chFPGA} for the CRS F-Engine. In Section~\ref{sec:validation}, we validate the digital signal processing (DSP) performance of \texttt{chFPGA} by injecting a wideband Gaussian noise source into a single CRS board. Finally, we summarize our results in Section~\ref{sec:conclusion}.

\subsection{Application: the Canadian Hydrogen Observatory and Radio transient Detector (CHORD)}

CHORD will be composed of a core array of $512$~six-meter dishes observing from $300$--$1500$~MHz at the Dominion Radio Astrophysical Observatory (DRAO), the Canadian federal observatory which hosts the Canadian Hydrogen Intensity Mapping Experiment (CHIME) near Penticton, British Columbia. In addition to the core array, two outrigger sites will be constructed at the Green Bank Observatory (GBO) in Green Bank, USA, and the Hat Creek Radio Observatory (HCRO) in Hat Creek, USA, separated from the core array by continental baselines which will enable CHORD to achieve precision localization of radio transients. The tremendous scientific achievements of CHIME (e.g., Refs.~\citenum{CHIMECrossCorr2022, CHIMEAutocorr2025, CHIMEFRBCat1_2021, CHIMEFRBCat2_2026}) have demonstrated the vast potential for driftscan interferometers for intensity mapping and radio transient studies. CHORD will build on the lessons learned from the CHIME design by prioritizing control over instrumental systematics while leveraging new technological advances since CHIME's inception. In pursuit of this goal, rapid progress is being achieved in commissioning the CHORD ``Pathfinder," a subset of the CHORD core array composed of $64$~dishes expected to be fully instrumented by the third quarter of 2026 \cite{Hoff2026Pathfinder}.

\subsection{The t0.technology Control and Readout System (CRS)}

\begin{figure} [ht]
\begin{center}
\begin{tabular}{c} 
\includegraphics[width = \linewidth]{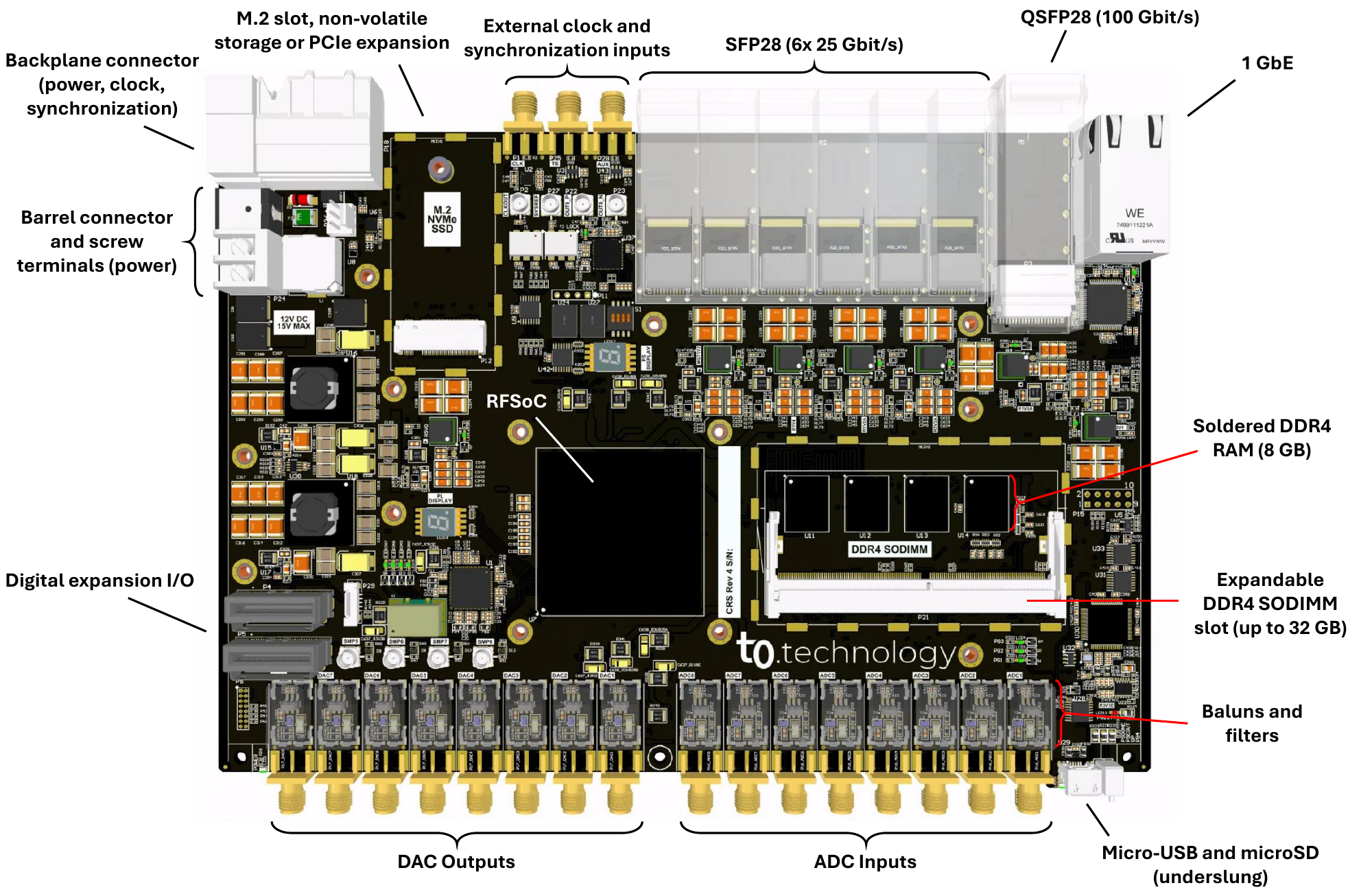}
\end{tabular}
\end{center}
\caption[example] 
{ \label{fig:example} 
A 3D rendering of a single CRS board. The RFSoC is located roughly at the center of the board, displayed as a black square with a white outline. The RF inputs and outputs are located at the bottom of the board through SMA connectors. High-speed data offload interfaces, external clock and time signal inputs, and backplane connections are located at the top of the board. Additional interfaces and peripherals are indicated.
}
\end{figure}

Advancements in RF and computing technologies has enabled radio correlators to achieve massive data throughput and real-time DSP for modern radio interferometers. The CHORD DSP system is designed around an F-X correlator architecture inspired by the CHIME DSP philosophy, with a field-programmable gate array (FPGA)-based F-Engine and a graphics processing unit (GPU)-based X-Engine. While the ICE system \cite{ICE} developed and used for CHIME was initially considered a candidate for the CHORD F-Engine platform, the t0.technology\footnote{\url{https://www.t0.technology/}} Control and Readout System (CRS)\cite{Montgomery2024CRS} was developed after CHORD was funded and subsequently  selected due to its expected improved performance for CHORD. 

The CRS is a multi-purpose FPGA-based microwave readout system offering high DSP performance and hardware scalability for applications in radio cosmology and astrophysics. The CRS uses a Generation 3 AMD Zynq Ultrascale+ ZU47DR RF-System-on-Chip (RFSoC)\footnote{\url{https://www.amd.com/en/products/adaptive-socs-and-fpgas/soc/zynq-ultrascale-plus-rfsoc.html}}, which offers improved DSP resources compared with the ICE system while also integrating a central processing unit (CPU), field programmable gate array (FPGA) resources, and RF data converters all within the same integrated circuit. A key advantage for CHORD of the analog-to-digital converters (ADC) offered by the RFSoC platform is that they are capable of directly sampling the $300$--$1500$~MHz CHORD science band, as opposed to CHIME's ICE system which hosts a slower $800$~MSPS ADC that would have required each RF signal to be split into multiple bands and separately digitized, demanding a more complex analog system that would have been less stable and more challenging to calibrate. The ADCs of the RFSoC were characterized in Ref.~\citenum{HendricksenMScThesis} in the context of CHORD, which demonstrated significantly improved ADC performance compared with the ICE system and concluded that the CRS is a suitable platform for the CHORD F-Engine. The ADCs are discussed further in Section~\ref{subsec:adc}.

Figure~\ref{fig:f_engine_diagram} displays a 3D rendering of the CRS with labeled components. The CRS is implemented on a $6$U~$160$~mm, 14-layer, Panasonic MEGTRON 6 PCB featuring SMA connections to eight~14-bit analog-to-digital converters (ADCs) with sampling rates up to $5$~GSPS and eight~14-bit digital-to-analog converters (DACs) with sampling rates up to $9.85$~GSPS; the DACs are not used for CHORD. A $1$~GbE RJ45 interface provides network access for sending control commands and slowly offloading periodic data snapshots. The CRS offers more substantial data offload capabilities through one~$100$~GbE QSFP28 and six~SFP28 cages facilitating total data offload speeds up to $250$~Gbit/s. 12~V power to the board is provided either through a barrel connector (for a $12$~V AC-DC adapter) or through the backplane connector (the backplane is introduced in the following section). Additional on-board hardware features include $8$~GB of soldered DDR4 RAM with an expandable DDR4 SODIMM slot, a slot for M.2 non-volatile data storage or PCIe expansion, a microSD slot, micro-USB connector, and digital expansion I/O interfaces. External reference clock and synchronization/timing (e.g., IRIG-B) signals can be provided to the CRS through three SMA connectors or through the backplane connector. The signals are critical for enabling intra- and inter-board synchronization, further described in the next section.

\subsection{The CRS Crate and Backplane}\label{sec:crate_and_backplane}

The CRS is designed with a philosophy of hardware scalability. It can be used as a standalone, single-board readout platform for small applications and laboratory testing, and for larger applications, it can be deployed in a $6$U-crates supporting up to 16 interconnected CRS boards, with their SFP and QSFP connectors enabling interconnection between multiple crates. The crates are equipped with a backplane, first introduced in Ref.~\citenum{Avelino2026CRSBackplane}, which is divided into four interconnected PCBs supporting $4$~CRS boards each; together, they implement the physical scaffolding enabling the scalability of the CRS platform by providing: (1) power distribution from the PSU; (2) external clock and timing signal distribution for synchronization of all RF data converters in a crate; and (3) interconnects allowing data shuffling between boards within a crate. The crates also include a software-controlled $3.3$~kW-rated power supply unit (PSU) as well as three computer fans seated directly beneath the CRS boards for cooling.

The CRS and its backplane were custom-designed to provide low differential phase noise between RF data converters in the crate as well as sub-sample phase reproducibility across power cycles. The low differential phase noise is achieved through a single low-jitter clock distribution chip on the backplane which feeds all CRS boards in a crate, with each board using a dual-phased-locked-loop (PLL) and a temperature-compensated crystal oscillator (TCXO) clock generation circuit designed to filter out external phase noise and favor the local oscillator for short-term stability and the external clock for longer-term synchronization. Path lengths are kept as short as possible to minimize thermal drifts, and clocking of the ADCs is done directly to avoid using the ADC's internal PLL, which may be adversely affected by the FPGA power system. 

Array-wide synchronization of all the ADCs with guaranteed phase reproducibility is achieved by: (1) resynchronizing the dual-PLL outputs of the PLLs to the external $10$~MHz reference provided by the backplane; (2) using a low-frequency clock to calibrate the ADC pipeline latencies; and (3) using a GNSS-generated IRIG-B timecode, pulse-per-second (PPS) signal, or centrally-generated pulse distributed by the backplane to each board to select a specific $10$~MHz reference clock edge on which data acquisition will be started on each board. Once thermally stabilized, the distribution of CRS boards can be synchronized to within a few milli-samples.

Each of the four backplane sub-modules in a CRS crate feature $25$~Gbit/s high-speed channels for data transmission between the four CRS boards within a sub-module, as well as four passive zQSFP+ (QSFP28-compatible) connectors which enable data transfer between other sub-modules. Together, these links can be used to implement a two-stage full-mesh network which can be leveraged for corner-turning of channelized data within a full crate. While the CHORD baseline plan for performing the corner-turn operation between the F- and X-Engines is to use the routing capabilities of  a $100$~GbE switch in the Pathfinder and a $400$~GbE switch for the full CHORD core array, it was necessary to ensure that the hardware and firmware of the CRS could support a backplane-based corner-turn operation to provide a risk-minimizing alternative for CHORD. Beyond its relevance for CHORD, that corner-turning capability has shown its utility in demonstrating a fully FPGA-based $32$-element (4-board) and a $64$-element (8-board) $N^2$ F-X correlator within a single CRS crate, which also confirmed the feasibility of a $128$-element (16-board) version.

\subsection{Infrastructure of the CRS F-Engine}\label{subsec:infrastructure}

\begin{figure} [h!]
\begin{center}
\begin{tabular}{c} 
\includegraphics[width = 0.7\linewidth]{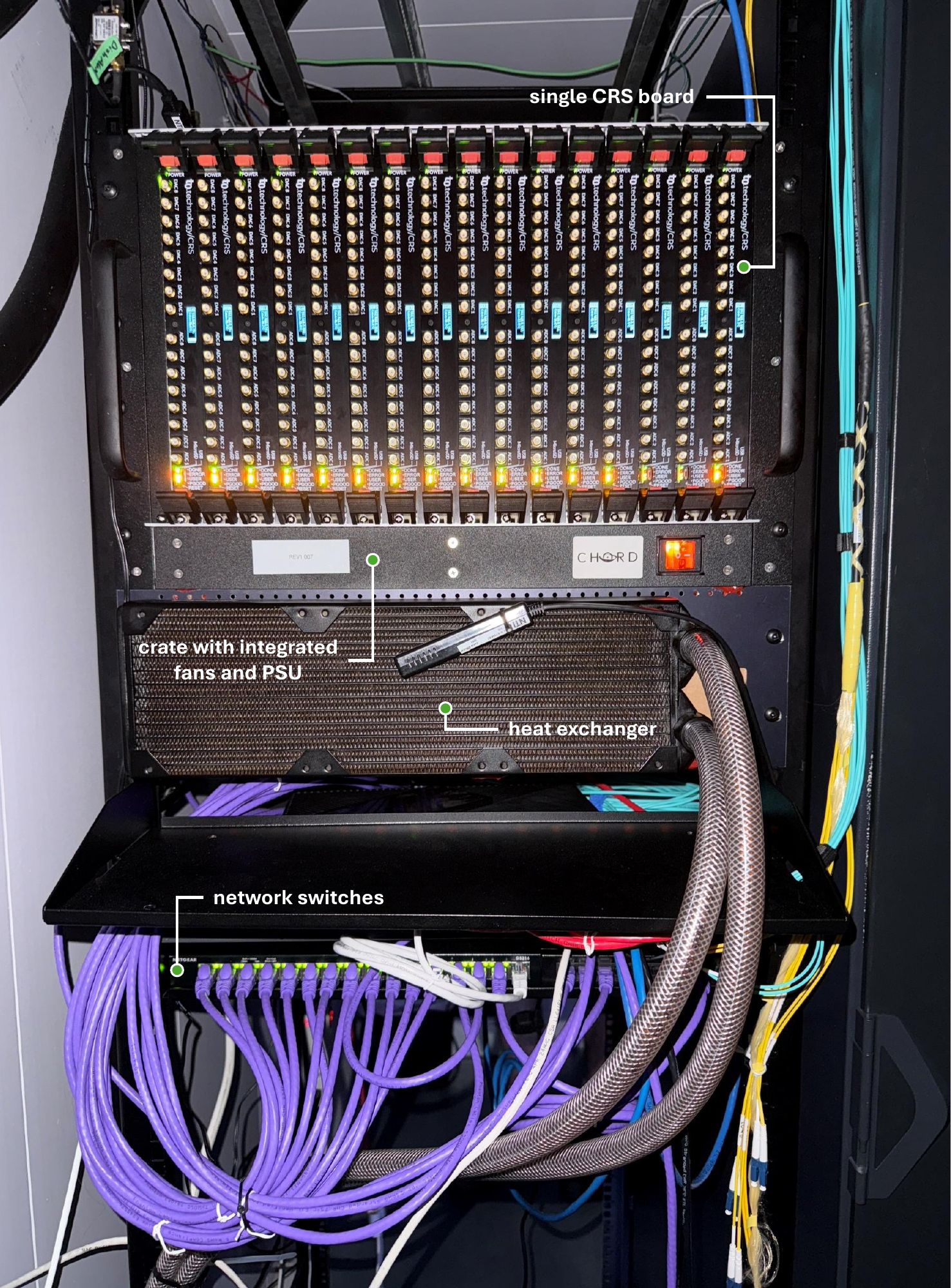}
\end{tabular}
\end{center}
\caption[example] 
{ \label{fig:f_engine_crate} 
One CRS F-Engine crate is shown in one CHORD receiver hut at DRAO. The CRS crate containing $16$~CRS boards is located at the top of the rack in the image, with each of the boards fully programmed and running the \texttt{chFPGA} firmware (described in Section~\ref{sec:firmware}). Immediately below the crate is the heat exchanger, which circulates a chilled coolant to extract heat from the room. Three temperature sensors are installed on the F-Engine: one at the air intake of the radiator, one above the CRS crate, and one behind the CRS boards (not pictured). Network switches are located below the drip tray, which connect to each of the boards at the back of the CRS crate and peripheral equipment. The GNSS receiver is located at the back of the rack on a separate shelf, which connects to the CRS backplane for clock and time signal distribution. The RF signal cables were removed for better clarity.
}
\end{figure}

The CHORD core array F-Engine will ultimately consist of $128$~CRS boards contained in $8$~crates. The crates are housed within two environmentally-shielded faraday enclosures (hereafter referred to as ``receiver huts") within the core array to minimize the radio-frequency interference (RFI) generated by digital electronics. Four CRS crates will be mounted in two 19" racks per receiver hut, along with peripheral support devices. RF signals are routed into the two receiver huts through aluminum and brass bulkhead panels. Figure~\ref{fig:f_engine_crate} displays an image of one CRS F-Engine crate, supporting $128$~RF signals from $64$~dual-polarized dishes. 

A Spectrum Instruments TM-4\footnote{\url{https://www.spectruminstruments.net/products/tm4/tm4.html}} GNSS receiver mounted in one receiver hut compute rack, along with its corresponding antenna on the roof (neither visible in Figure~\ref{fig:f_engine_crate}), distributes a GNSS-disciplined $10$~MHz clock and IRIG-B to the F-Engine crates for synchronization of the array as described in the previous section. An industrial water chiller housed next to each receiver hut provides cooling for the heat generated by the F-Engine by cycling a liquid coolant through a liquid to air heat exchange radiator (located immediately below the F-Engine crate in Figure~\ref{fig:f_engine_crate}); the fans of the CRS crate pull warm ambient air through the radiator, removing heat and passing the chilled air over the CRS boards. An external computer running the F-Engine software \texttt{pychfpga} (introduced in subsection~\ref{subsec:pychfpga}) enables remote control, data collection, and firmware, software, and hardware metric monitoring of the CHORD F-Engine. The X-Engine implementation for the full CHORD core array will be housed in a physically separate RF-shielded enclosure $\sim 150$~m from the F-Engine receiver huts, and data transmission to the X-Engine will be facilitated by single-mode fiber transceivers and a $400$~GbE switch. For the validation described in this paper, the X-Engine is not used.

\section{CRS F-ENGINE FIRMWARE AND SOFTWARE}\label{sec:firmware}

\begin{figure} [ht]
\begin{center}
\begin{tabular}{c} 
\includegraphics[width = \linewidth]{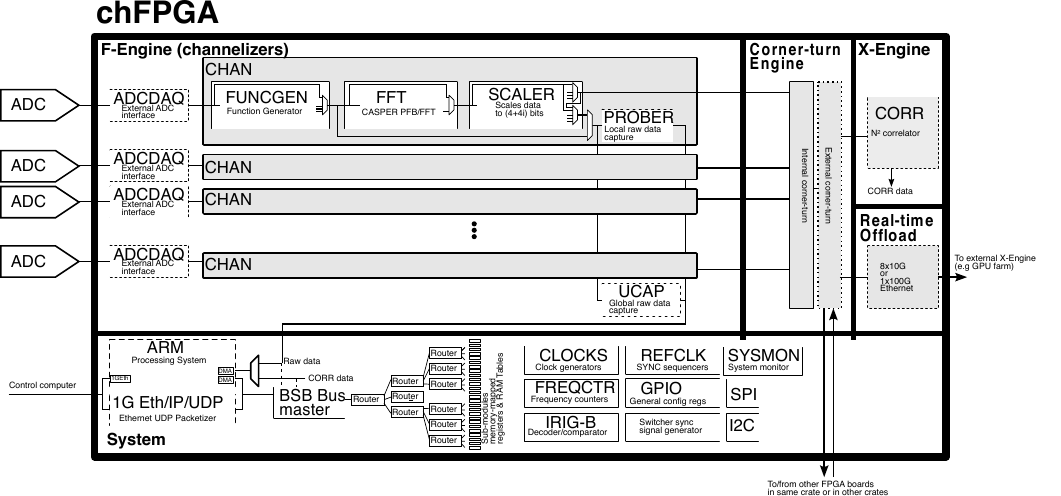}
\end{tabular}
\end{center}
\caption[example] 
{ \label{fig:f_engine_diagram} 
A generalized block diagram of the \texttt{chFPGA} digital signal processing firmware, with the data flow read from left to right. Eight~ADCs digitize the analog signals, which are processed by the ADCDAQ module before entering the channelizer. The channelizer signal processing includes (1) a function generator, which can optionally replace the ADC timestream data with repeated arbitrary waveforms; (2) a CASPER-based PFB/FFT, which performs the channelization; (3) the Scaler, which re-quantizes the channelized data to a much smaller bit-depth; and (4) the \texttt{UCAP} data frame capture engine, which can periodically capture bursts of data at different stages of the digitization and channelization signal processing. The re-quantized FFT data is then processed by a corner-turn engine for input signals on a single board and/or with additional boards depending on the firmware configuration. There are two separate firmware versions for the final stage of signal processing: (1) a packetizer engine to efficiently assemble the re-quantized FFT data into Ethernet packets, which are transmitted to an external X-Engine using a $100$~GbE transceiver core, or (2) \texttt{UCORR}, a fully FPGA-based X-Engine for single-board $(N = 8)$, 4-board $(N = 32)$, and 8-board $(N = 64)$ $N^2$ correlator options. The ARM processor of the RFSoC is used for managing the CRS and provides an interface to the firmware signal processing and system modules.
}
\end{figure}

The CRS F-Engine signal processing is based on \texttt{chFPGA}, an open-source, multi-platform FPGA firmware for radio interferometers developed in partnership by McGill University and t0.technology \cite{chFPGA}. \texttt{chFPGA} offers an array of ADC data acquisition modules connected to downstream channelization modules which condition the data, perform a channelization, and scale the results to $(4 + 4i)$~bits using programmable per-frequency gains. The following stage performs an on-board (and optionally off-board) corner-turn operation, whose output can be offloaded to an external X-Engine through a $100$~GbE link or can alternatively be sent to an internal $N^2$ X-Engine, the latter of which has proven essential for prototyping and firmware validation. Therefore, while \texttt{chFPGA} was designed primarily with a focus on CHORD DSP requirements, it leverages the same general architecture for multiple applications, including fully FPGA-based F-X correlators on a single board or a full crate, by swapping these final stages of the signal flow in separate bitstream files. The on-board ARM processor provides system management and exposes a network-based interface to the firmware’s internal configuration and status registers. The signal processing flow of \texttt{chFPGA} is described in further detail in subsections~\ref{subsec:adc}--\ref{subsec:100GbE_engine}, and a diagram representing key stages of the firmware is presented in Figure~\ref{fig:f_engine_diagram}. 

The \texttt{pychfpga} package provides a Python-based interface for self-discovery and keeping track of every board and crate on the network and orchestrating array initialization, synchronization, monitoring, and general day-to-day telescope system operations \cite{pychfpga}. \texttt{pychfpga} is further described in subsection \ref{subsec:pychfpga}. 

A summary of key CHORD F-Engine quantities and their associated symbols are presented in Table \ref{tab:quantities}, which are introduced throughout this section.

\begin{table}[ht]
\caption{A summary of key quantities describing the CHORD implementation of the CRS F-Engine and their associated symbols (where applicable), for both a single CRS board and the full CHORD core array.} 
\label{tab:quantities}
\begin{center}       
\begin{tabular}{|l|l|l|} 
\hline
\rule[-1ex]{0pt}{3.5ex}  \textbf{Quantity} & \textbf{Symbol} & \textbf{Value} \\
\hline
\rule[-1ex]{0pt}{3.5ex}  ADC inputs per CRS board & $N_\text{inputs}$ & $8$ \\
\hline
\rule[-1ex]{0pt}{3.5ex}  Boards per crate & - & $16$ \\
\hline
\rule[-1ex]{0pt}{3.5ex}  Total F-Engine crates for CHORD& $N_\text{crates}$ & $8$ \\
\hline
\rule[-1ex]{0pt}{3.5ex}  Total CRS boards& $N_\text{boards}$ = $N_\text{inputs} \times N_\text{crates}$ & $128$ \\
\hline
\rule[-1ex]{0pt}{3.5ex}  ADC sampling rate & $f_\text{s}$ & 3.2 GSPS \\
\hline
\rule[-1ex]{0pt}{3.5ex}  ADC resolution & $\Delta$ & 14 bits \\
\hline
\rule[-1ex]{0pt}{3.5ex}  ADC samples per frame & $M$ & $2^{14} = 16,384$ samples \\
\hline
\rule[-1ex]{0pt}{3.5ex}  Frame length & $t_\text{frame} = M / f_\text{s}$ & $5.12$~$\mu$s \\
\hline
\rule[-1ex]{0pt}{3.5ex} F-Engine total digitized data rate & $\Delta \times f_\text{s} \times  N_\text{inputs} \times N_\text{boards}$ & $\sim 46$ Tbit/s \\
\hline
\rule[-1ex]{0pt}{3.5ex}  FPGA signal processing clock speed & $f_\text{s} / N_\text{samples/clock}$ & $400$~MHz \\
\hline
\rule[-1ex]{0pt}{3.5ex}  Samples processed per clock & $N_\text{samples/clock}$ & $8$~samples \\
\hline
\rule[-1ex]{0pt}{3.5ex}  PFB Architecture & - & CASPER \texttt{pfb\_fir\_real} \\
\hline
\rule[-1ex]{0pt}{3.5ex}  PFB window type & - & Hamming \\
\hline
\rule[-1ex]{0pt}{3.5ex}  Number of PFB taps & $N_{\text{taps}}$ & $4$ taps \\
\hline
\rule[-1ex]{0pt}{3.5ex}  PFB coefficient resolution & - & $18$ bits \\
\hline
\rule[-1ex]{0pt}{3.5ex}  Total number of frequency bins & $N_{\text{bins}} = M/2$ & $8,192$ bins \\
\hline
\rule[-1ex]{0pt}{3.5ex}  Frequency resolution & $f_\text{res} = f_\text{s} / M$ & $195.3125$ kHz \\
\hline
\rule[-1ex]{0pt}{3.5ex}  FFT coefficient resolution & - & $(18+18i)$~bits \\
\hline
\rule[-1ex]{0pt}{3.5ex}  FFT output resolution & - & $(32+32i)$~bits \\
\hline
\rule[-1ex]{0pt}{3.5ex}  FFT architecture & - & CASPER \texttt{fft\_wideband\_real} \\
\hline
\rule[-1ex]{0pt}{3.5ex}  FFT bitgrowth option enabled & - & Yes \\
\hline
\rule[-1ex]{0pt}{3.5ex}  ``Scaler" re-quantized FFT resolution & - & $(4+4i)$ bits \\
\hline
\rule[-1ex]{0pt}{3.5ex}  Frequency bins transmitted to X-Engine & - & $6,144$~bins ($300$--$1500$~MHz) \\
\hline
\rule[-1ex]{0pt}{3.5ex}  Output interface to X-Engine & - & $100$~GbE (QSFP28) \\
\hline
\rule[-1ex]{0pt}{3.5ex}  Output data rate (single board) & - & $76.80625$~Gbit/s ($300$--$1500$~MHz) \\
\hline
\rule[-1ex]{0pt}{3.5ex}  Total output data rate ($128$~boards) & - & $9.8312$~Tbit/s \\
\hline
\end{tabular}
\end{center}
\end{table}

\subsection{ADC Acquisition Subsystem}\label{subsec:adc}

The implementation of \texttt{chFPGA} for CHORD uses all $N_\text{inputs} = 8$~ADC inputs provided by the FPGA and exposed by the CRS, and digitizes analog signals at $f_\text{s} = 3.2$~GSPS with the full $\Delta = 14$-bit resolution. The ADCs use interleaved sampling, whereby four sub-ADC ``cores" individually sample the same signal at $0.8$~GSPS with a phase offset and combine their signals to attain the full $3.2$~GSPS sampling rate. The interleaving of the sub-ADC cores, whose relative characteristics are sensitive to temperature fluctuations, can introduce artifacts into the spectrum which manifest as spurious tones and intermodulation distortion (IMD) products. The sub-ADC cores are therefore calibrated for their relative gain, phase, and DC offset using the internal real-time calibration system provided by the ADC to suppress these artifacts \cite{RFSoCADCGuide}. A ``foreground" calibration is applied upon power-up of the ADCs, and a ``background" calibration may be optionally applied during normal ADC operations to continuously update the calibration coefficients. The coefficients can optionally be ``frozen" and thus not updated in real-time. The background calibration requires an input signal level $\geq -40$~dBFS. For the measurements introduced in this paper, the background calibration is always enabled while conducting measurements, which is preferred in order to maintain cohesive sampling across the sub-ADC cores. \texttt{chFPGA} exposes registers to monitor and enable or freeze the background calibration of the ADCs, as well as monitor and reset status flags, such as ADC overflows. 

The firmware divides the ADC data stream into ``frames" of $M = 2^{14} = 16,384$~samples of total length $t_\text{frame} = 5.12$~$\mu$s, which ultimately defines the number of frequency bins and thus their resolution (described in the following section). The sampling rate of $3.2$~GSPS is slightly higher than the $3$~GSPS required to Nyquist-sample the CHORD $300$--$1500$~MHz science band, resulting in a first Nyquist zone corresponding to DC--$1600$~MHz. The extra $100$~MHz of bandwidth gained from the faster sampling rate provides a guard band that greatly relaxes the roll-off stability, reproducibility, and steepness requirements of the analog anti-aliasing low-pass filters (LPF), which also corresponds with smaller-amplitude and lower-frequency ripples in the filter passband. A slow LPF roll-off is thus preferred for CHORD intensity mapping objectives where smoother passbands minimize foreground leakage into target science modes \cite{LiuShaw21cmReview}. Further, digitizing a guard band to accommodate a slower LPF roll-off is advantageous given that a steep LPF roll-off introduces higher temperature susceptibility and lower reproducibility between different filter instances.

The performance of the ADCs in the context of CHORD was characterized in Ref.~\citenum{HendricksenMScThesis}, which concluded that the ADCs as implemented on the CRS platform demonstrated that the performance exceeds the requirements for CHORD. The noise spectral density (NSD) was measured to be $-154$~dBFS/Hz with an effective number of bits (ENOB) of $10$~bits. The spurious-free dynamic range (SFDR) was measured to be $< -78.5$~dBFS; the SFDR includes IMD products introduced by the interleaving of the sub-ADC cores, demonstrating that these products are sufficiently suppressed with continuous background calibration active. The crosstalk between ADC channels on the same CRS board was found to be $< - 72$~dBFS, a significant improvement in comparison with that of the ICE platform, with a specification of $< -46$~dBFS for channels on the same ADC daughter mezzanine. Overall, the ADC performance was found to be consistent with the datasheet and either improved or consistent compared to the ICE system's ADCs, while offering the benefit of sampling the entire CHORD band uniformly without the complexity and systematics associated with sub-sampling the band.

\subsection{Channelizer Subsystem}

This section describes the ``channelizer" subsystem, whose sub-modules separates the incoming digitized timestream data from the ADC acquisition subsystem into individual frequency channels.

\subsubsection{Function Generator}

The first signal processing block of the channelizer is the function generator. In the default pass-through mode, the function generator passes ADC data to the channelizer with no modification, while tracking statistics such as ADC overflows to characterize the observing conditions. Optionally, the function generator can scale or clip the data as it passes through, or replace the ADC data stream entirely with custom test patterns.

The function generator, as its name implies, can also replace the ADC data stream with various waveforms and test patterns that are useful in testing the downstream signal processing pipeline modules. In one mode, the user can program a full frame of any arbitrary waveform and have it played back repeatedly. The data can also include custom FFT spectra can that can be fed downstream by bypassing the PFB/FFT (further detailed in the following section). Other useful modes include a pseudorandom noise (PRN) generator with user-defined seeds, user-defined simulated overflow patterns, or frames filled with values corresponding to the sample and frame number, which can be useful for diagnosing data routing all the way to the X-Engine.

\subsubsection{CASPER-Based PFB/FFT}
The channelization of the timestream data is performed using a polyphase filter bank (PFB) and a fast Fourier transform (FFT) developed by the Collaboration for Astronomy Signal Processing and Electronics Research (CASPER) and available open-source\cite{CASPER}.

The channelization begins with a critically-sampled PFB generated with the CASPER \texttt{pfb\_fir\_real} block, whose purpose is to reduce the channel leakage inherent to the boxcar window function of an FFT. An $N_\text{taps} = 4$-tap Hamming window was selected with $18$-bit coefficient resolution to strike a balance between the resource utilization of memory and DSP units, considerations of upchannelization requirements for the radio transient backends of CHORD, and minimization of window-induced side-lobe leakage. The PFB promotes the 14-bit depth of the ADC timestream into an output bit-depth of $18$~bits.

After passing through the PFB, the $M = 16,384$~samples of a single frame are Fourier-transformed into $N_\text{bins} = 8,192$~positive frequency bins with resolution $f_\text{res} = 195.3125$~kHz. Eight ADC samples are channelized to produce $4$~real frequency bins per clock per ADC input. The FFT uses the \texttt{fft\_wideband\_real} CASPER block, which implements a decimation-in-frequency FFT with a biplex (direct) architecture for the first $11$~stages (final $3$~stages) for efficient FPGA resource utilization. The FFT coefficients have a bit-depth of $(18+ 18i)$~bits. The bit-growth option of the FFT is enabled, wherein the bit-depth of data passing through the FFT stages grows by one bit in its real and imaginary components at each stage, preventing overflow at the expense of additional FPGA resources. The FFT therefore begins with a bit-depth of $(18 + 18i)$~bits at the first stage and grows to $(32 + 32i)$~bits by its final stage. To compensate for the resource consumption demanded by enabling the bit-growth option, the option to reorder the frequency bins was left disabled, which makes the FFT return the frequency bins in its natural decimation-in-frequency order as opposed to an ascending order. This requires less resources in the FPGA and is later compensated for in downstream software.

\subsubsection{Scaler Post-FFT Re-Quantization}\label{sec:scaler}

The tremendous amount of data generated by the FFTs on a single CRS board ($819.2$~Gbit/s in total for all $8$~input channels) would be impractical to transmit from the F-Engine to the external X-Engine, nor could the X-Engine process that data in real-time. Fortunately, as successfully demonstrated on similar F-X-correlator-based telescopes like CHIME, the channelized data can be re-quantized to a lower bit-width (and therefore lower data rate) prior to data transmission to the X-Engine while minimizing quantization bias\cite{CHIMEOverview}. This re-quantization is the role of the Scaler sub-module. 

The FFT re-quantization is well-motivated by considering the behavior of the received signals, which includes emission from the cosmic microwave background (CMB), diffuse and discrete astronomical sources, blackbody emission from the ground, noise from the electronic receiver, and RFI. The signal of interest from the sky can generally be modeled as a zero-mean Gaussian distribution (though temperature-induced fluctuations in the system gain can complicate this model) \cite{EssentialRadioAstronomy}, and therefore typically has a low dynamic range whose power is spread across the bandwidth. In contrast, RFI often concentrates power in a small number of frequency bins. As such, significant bit-growth in the FFT occurs primarily for RFI-contaminated frequency bins during channelization, meaning that the majority of the dynamic range of the FFT is used for signals which are not useful (though it is crucial to provide this wide dynamic range to prevent possible FFT overflows caused by RFI). It is therefore possible to re-quantize the data to a much lower bit-depth without adding significant quantization noise to the sky signals, noting that maintaining the full bit-depth of RFI-contaminated frequency bins is not necessary since they will ultimately be discarded.

Given the above, we can model the sky signal in non-RFI-contaminated frequency bins as a complex, zero-mean, circularly symmetric Gaussian whose width does not change significantly on the timescale of a single integration. This is generally a valid assumption, since typical integration times in radio astronomy are of order $10$~s, while temperature-induced gain fluctuations vary on timescales of minutes to hours. Therefore, the optimal quantization levels for a correlator can be pre-determined through simulation to inform how to best tune the real implementation of the post-FFT re-quantization \cite{CorrelatorQuantizationBiasJuan2018}. As in the CHIME case, the $(32+32i)$-bit FFT data is quantized to $(4+4i)$~bits and clipped to $-7$ to $+7$ in both the real and imaginary parts such that the mean value is zero and not biased towards the negative side. The quantization is performed by multiplying the FFT data by a unique ``digital gain" for each frequency bin, resulting in a $(48+48i)$-bit number of which only the upper $(4+4i)$~most significant bits are passed to the next signal processing stage.

The signal levels of the quantized $(4+4i)$-bit data for each frequency bin are tuned to avoid signal levels (defined here as the average magnitude of the re-quantized FFT data) which are (1) too low, where quantization noise begins to dominate, and (2) signal levels which are too high, where the values begin to saturate. We adopt a target $(4 + 4i)$-bit magnitude within the range recommended by Ref.~\citenum{CorrelatorQuantizationBiasJuan2018} and used by CHIME with its $(4+4i)$-bit FFT re-quantization firmware. Since all frequency bins are assumed to be described by the same Gaussian process, the same optimal quantization level applies for all frequency bins. In practice, the digital gains therefore effectively flatten the average magnitude to the same target value, which is passed to the X-Engine for correlation.

We determine the optimal digital gains by collecting many FFT frames in Python, typically $\mathcal{O}(100$--$1000)$, averaging them over all frames for each input, and directly calculating the digital gains by dividing the target by the integrated FFT data. The direct calculation (as opposed to CHIME's iterative digital gain optimization process), is enabled by a data capture mode which outputs the full $(32 + 32i)$-bit FFT outputs by interleaving odd and even bins in alternating data frames. This direct calculation significantly reduces the digital gain computation time, from several minutes with CHIME to $\sim 10$~s with CHORD. 

These quantized digital gains are stored in HDF5 files for offline calibration of correlated data, which is necessary to convert visibilities into values which are proportional to power by re-introducing the bandpass structure removed by the digital gains. Notably, the data capture path is independent of the ``science" data path (i.e., re-quantized FFT data flow sent to the X-Engine), meaning that digital gains can in principle be updated on a periodic basis without affecting telescope operations. Further, the Scaler memory banks are large enough to store two separate sets of digital gains, offering the possibility to switch between gain sets to accommodate variations in signal levels (though any such switching would need to be well-documented with metadata for informing science backends). The Scaler subsystem offers various additional data capture modes (e.g., recovering the intermediate $(48 + 48i)$-bit quantity for debugging purposes), including a ``statistics" module which can track the number of post-re-quantization overflows over programmable periods of time on the non-science data capture path, either for a single bin or all frequency bins.

\subsubsection{\texttt{UCAP} Data Frame Capture Engine}\label{subsec:ucap}

\texttt{UCAP} is a flexible data capture engine which taps into the channelizer's function generator and Scaler outputs and can periodically capture bursts of frames, serialize them, and slowly stream them back to the host computer for continuous monitoring the real or simulated ADC timestreams, the FFT output, and Scaler output. It is a parallel data capture stream, such that it can be operated simultaneously with the standard ``science" data offload path. \texttt{UCAP} takes advantage of the high-density ``UltraRAM" (URAM) memory available on the Ultrascale+ family of FPGA. The captured data is transmitted through user datagram protocol (UDP) packets over the $1$~GbE link. The user can set the capture rate to any desired value as long as the resulting stream does not saturate the link. A total of $16$~frames are captured and transmitted in a single burst, with the user able to chose whether to capture  $16$~contiguous frames ($16 \times M = 262,144$~samples, or $81.92$~$\mu$s) for a single input signal, or $2$~contiguous frames for each of the $8$~input signals ($2 \times M = 32,768$~ samples, or $10.24$~$\mu$s), as well as the in-between combinations. As mentioned in the previous section, the data capture rate and configuration can be changed during operation without affecting the ``science" data flow of re-quantized FFT data transmitted to the X-Engine.

\subsection{Corner-Turn Engine}\label{subsec:corner_turn}

\texttt{chFPGA} implements a simple internal corner-turn operation between input channels on a single CRS board. This corner-turn is implemented only by internal FPGA routing, and feeds the packet assembler and transmitter (described in the following section). \texttt{chFPGA} supports additional corner-turning capability for non-CHORD applications in which data is exchanged between  boards on the same 4-slot backplane using three $25$~Gbit/s links (level-2 corner-turning), and between boards on other 4-slot backplanes using three more $25$~Gbit/s links (level-3 corner-turning).

Following this point, the data can pass through one of two options for additional signal processing depending on the application and firmware version being used: (1) a packet assembler and $100$~GbE data transmission link for sending FFT data to the X-Engine (``CHORD firmware mode"), or (2) a fully FPGA-based $N^2$ correlator engine. These two firmware versions are described in the next two sections.

\subsection{\texttt{UPACK} Packet Assembler and CGE 100 GbE Data streamer}\label{subsec:100GbE_engine}

The \texttt{UPACK} module takes the internally corner-turned data and stores it in the dense URAM blocks in the FPGA. Specifically, $16$~consecutive frames of $8,192$ frequency bins from all $8$~input channels are stored in this fashion. Once a full data set is collected, a readout state machine reads the data bin-per-bin, $16$~frames at a time, and streams the data to the $100$~GbE transmitter while another set of frames is being assembled. The readout of the bins is orchestrated by a programmable readout table that lists which sets of bins are assembled into packets and where they should be sent. Some URAM locations that would normally hold data for a subset of frequency bins are user-programmable and used to store the Ethernet/IP/UDP packet headers that are placed at the beginning of each packet. This allows the user to send any set of bins to any IP address in any order, granting a high degree of control over the routing of specific frequency bins to destination X-Engine nodes. In addition to the routing information, the header contains codes to identify the specific CRS board and frequency bins corresponding to the payload, as well as a $64$-bit timestamp recorded as the frame number since frame~$0$, among other metadata.

The data generated by \texttt{UPACK} is fed to a built-in $100$~GbE hardware complex multiply–accumulate (CMAC) core (the ``CGE" module) to be transmitted as four $25$~Gbit/s data links fed to the CRS QSFP cage. The transmitter's Reed-Solomon Forward Error Correction (RS-FEC) can be enabled or disabled in software, depending on what type of QSFP transceiver and media (copper, 4x multi-mode fiber, or single-mode fiber) is used.

The total data rate for a single CRS board following re-quantization of the FFT is $102.4$~Gbit/s if all $8,192$ frequency bins are transmitted, which is just above the maximum data rate of the $100$~GbE link (though the in-practice maximum is likely lower), without including the packet headers. However, since only a subset of these bins is used for CHORD science analysis, the bin readout tables are programmed to transmit the $6,144$~bins corresponding to the CHORD $300$--$1500$~MHz science band to the X-Engine through $128$~packets containing $48$~bins each, which minimizes the data rate sufficiently so as not to saturate the $100$~GbE link. As mentioned earlier, some of the remaining URAM memory gained by dropping some of the frequency bins are used to for storing packet header information. The total output data rate of a single CRS board is therefore approximately $76.80625$~Gbit/s, or a total of $9.8312$~Tbit/s transmitted from F- to X-Engine for all $128$~boards for the CHORD core array implementation. 

As described in Section~\ref{subsec:infrastructure}, CHORD plans to use a network switch to route and corner-turn packets containing each set of frequency bins to specific GPU nodes. If all boards are programmed to send the same frequency bins to the same destination node, the GPU can reassemble the data from multiple CRS boards and thus complete the corner-turn operation before correlating the channelized data.

\subsection{\texttt{UCORR} FPGA-Based X-Engine}\label{subsec:ucorr}

The output of the corner-turn engine can be connected to the \texttt{UCORR} module, a fully FPGA-based X-Engine which supports single-board (the ``Pocket Correlator") and multi-board $N^2$ correlators. \texttt{UCORR} cannot be included alongside the \texttt{UPACK} packet assembler, as they share the same URAM resources. Users choose which mode to operate in using different bitstream files. \texttt{UCORR} takes the corner-turned (4+4i)-bit re-quantized FFT outputs from all the channelizers and computes all visibility products in real-time using a highly optimized 2-DSP kernel, and integrates those products within its internal URAM and standard Block RAM (BRAM) banks. Once the on-board integration period is complete, the products are slowly streamed to a control machine using the $1$~GbE link while the next integration is being processed. 

The user can integrate up to $65,536$ frames ($\sim 336$~ms) of FFT data with the Pocket Correlator firmware. The integrated values are internally stored as $(18 + 18i)$-bit values. The integrated data bit-depth imposes an upper limit on the integration period; long integration periods will decrease the output data rates, but will increase the possibility that the integrated values will saturate the $(18 + 18i)$-bit values. On the other hand, the Ethernet link also places a data rate limitation on the integration period, as a shorter integration period causes the data rate to increase and potentially saturate the $1$~GbE link. In practice, $2^{14}$ frames ($\sim 84$~ms) integrated with \texttt{UCORR} combined with FFT data re-quantized according to the target described in Section \ref{sec:scaler} strikes an optimal balance between the two. Deeper integrations are achieved by recovering multiple firmware-integrated frames and performing further integration in software using \texttt{pychfpga}. Standard integration times of $\mathcal{O}(10$~s) employed in radio cosmology are therefore easily achieved with the Pocket Correlator. 

The Pocket Correlator is thus an excellent tool not only for benchtop applications with modest computing resources for validating the performance of the CRS platform and the radio firmware, but it can also be used for interferometers with $8$ or fewer input signals. In particular, the Pocket Correlator was demonstrated with success on the Deep Dish Development Array (D3A) \cite{Islam2020D3A, Islam2022D3A6}, a three-element interferometer at DRAO used for developing the key CHORD analog and digital signal processing technologies. 

We further note that \texttt{UCORR} is fully scalable and has been used to implement $N = 32$ and $N = 64$ multi-board correlators in conjunction with level-2 or level-3 corner-turning logic. However, the total bandwidth through the $1$~GbE links is conserved, even as we increase the number of spatial products. Therefore, for the $N = 32$ and $N = 64$ configurations, we must either cycle through a subset of the frequency bins, or continuously stream data products for a subset of the total number of frequency bins. We note that in principle, the data could be streamed to a control machine over a $10$~GbE link using one of the SFP28 cages, but \texttt{UCORR} is fundamentally limited by DSP and memory resources in the FPGA, and a faster Ethernet link would thus not enable the instantaneous recovery of more frequency bins.

\subsection{\texttt{pychfpga} Python-Based Control and Monitoring}\label{subsec:pychfpga}

\texttt{pychfpga} is an open-source Python package designed for automating the control and monitoring of the F-Engine during telescope operations. Initially developed for the ICE system\cite{ICE}, it has since been upgraded to support all the platforms on which \texttt{chFPGA} can run, including the AMD ZCU111 Evaluation Kit\footnote{\url{https://www.amd.com/en/products/adaptive-socs-and-fpgas/evaluation-boards/zcu111.html}} (used during the early stages of CHORD F-Engine development) and the CRS. Multiple firmware configurations (e.g., the $100$~GbE packet assembler, the FPGA-based X-Engine, etc.) are supported for each of those platforms.

\texttt{pychfpga} provides low-level objects that represent each module in the firmware (the function generator, PFB/FFT, Scaler, \texttt{UCORR}, etc). These objects allow seamless access to the module's memory-mapped registers and memory blocks through simple attributes and standardized methods. The objects also provide a higher-level application programming interface (API) that hides the implementation specifics of the module, such as enabling the FFT to be bypassed, injection of function generator waveforms, writing Scaler gains to BRAM, monitoring RFSoC voltage and temperatures, etc. A global \texttt{chFPGA} object (here referring to the software interface) ties together all the object instances representing sub-modules together to present a unified interface to an FPGA. The \texttt{chFPGA} object establishes network communication with the RFSoC, reads the configuration registers and creates all the objects that correspond to that configuration, and provides its own set of higher-level API to setup chip-wide signal processing functionalities.

\texttt{FPGAArray} is a higher-level object for managing multiple CRS boards simultaneously. It is responsible for discovering the boards on the network (over mDNS or a custom beacon protocol), instantiating the motherboard object for the corresponding platform, loading the RFSoC with the appropriate firmware, and instantiating the corresponding Python object for that firmware. Once initialization is complete, it provides array-wide functions, such as signal processing configuration and array synchronization that enables configuration of all boards in an array through a single object. Further, these functions are executed asynchronously using Python coroutines such that all boards can be operated in parallel, which significantly reduces the initialization time of large arrays like CHORD.

 \texttt{FPGAMaster} is the top-level F-Engine class that operates the F-Engine during observations. At startup, it loads a YAML file that specifies all key configuration parameters of the F-Engine, such as the CRS board serial numbers (or instead, their IP addresses) to include in the array, the various processing modes, synchronization methods, gain calibration parameters, data capture speed and output folders, logging level, etc.  Once the boards are initialized and synchronized, \texttt{FPGAMaster} computes and initializes the Scaler module's digital gains for re-quantization of channelized data, and saves the digital gains to an HDF5 file for offline analysis. \texttt{FPGAMaster} also launches a HTTP/REST server that allows the user to initiate commands during an observation run (e.g., get system status, switch between two memory banks of digital gains, recompute digital gains, monitor the timing of frames, etc). It also periodically monitors the state of every hardware and firmware subsystem, and publishes those through one REST endpoint to be scraped by the metrics and monitoring software Prometheus\footnote{\url{https://prometheus.io/}}. The metrics are ultimately published to the monitoring user interface Grafana\footnote{\url{https://grafana.com/}} to provide an overview of the system health of the F-Engine (along with other CHORD hardware) through a graphical interface.

The \texttt{RawAcqReceiver} class runs a separate HTTP/REST server which operates in concert with \texttt{FPGAMaster} during regular telescope operations for collecting raw intermediate data products from the F-Engine, such as ADC timestream data for monitoring input signal power and the $(32 + 32i)$-bit FFT data for calculation of the digital gains, which are saved to HDF5 files. \texttt{RawAcqReceiver} can also receive and save data from the on-board X-Engine if using this version of the firmware.

\section{CRS F-ENGINE DSP VALIDATION}\label{sec:validation}

\begin{figure} [h!]
\begin{center}
\begin{tabular}{c} 
\includegraphics[width = 0.9\linewidth]{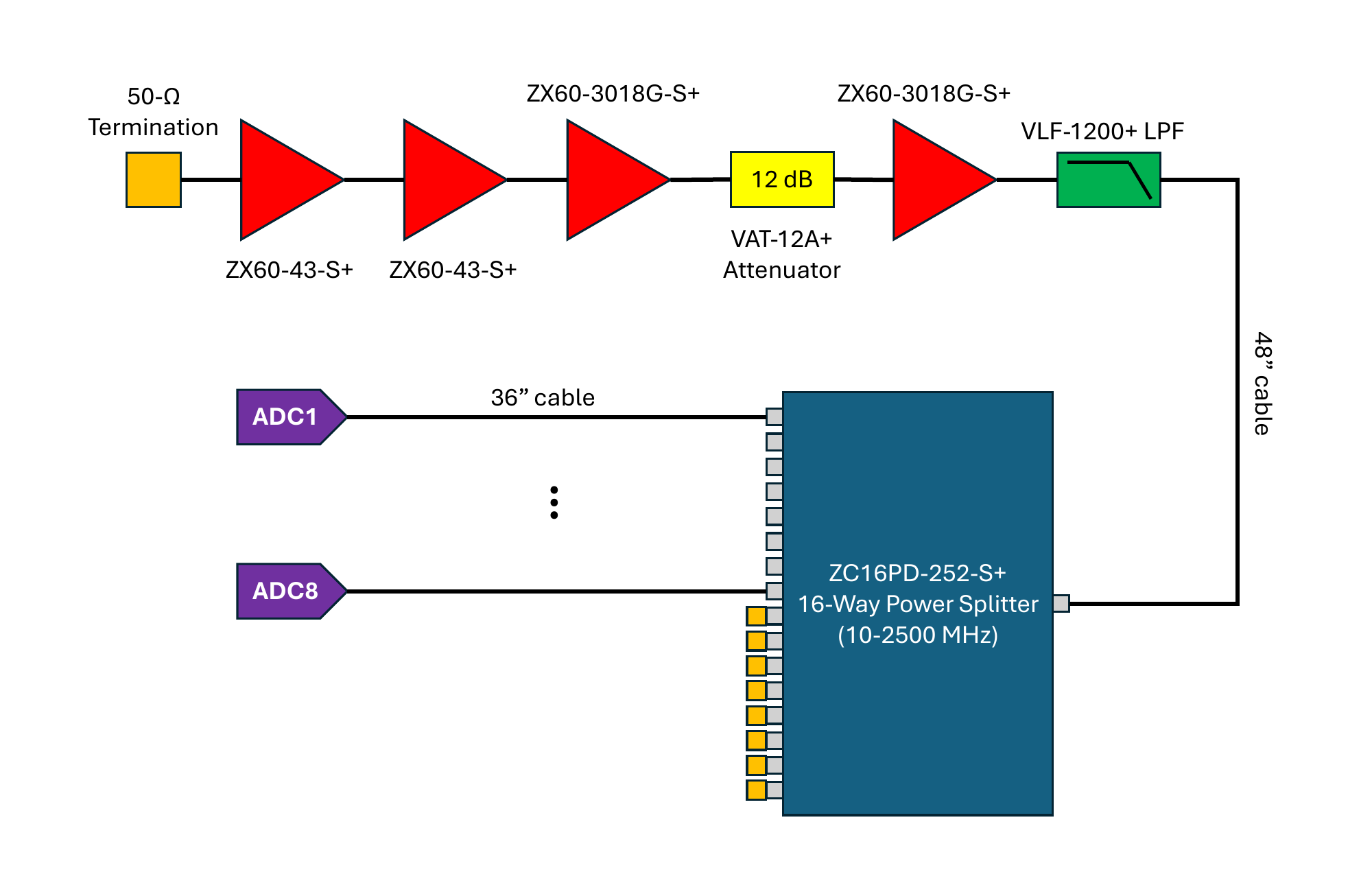}\end{tabular}
\end{center}
\caption[example] 
{ \label{fig:signal_chain} 
A block diagram of the signal chain used to produce a wideband, zero-mean Gaussian signal using an amplified and filtered $50$-$\Omega$ termination. All the analog components shown in the block diagram are Mini-Circuits devices. The $50$-$\Omega$ termination is first amplified through two ZX60-43-S+ and one ZX60-30186-S+ amplifiers, followed by a $12$~dB attenuator to reduce the signal power prior to entering an additional ZX60-30186-S+ amplifier. The signal is then filtered with an anti-aliasing VLF-1200+ LPF, which has its $3$~dB point at $1530$~MHz, and connected through a 48" cable to a ZC16PD-252-S+ 16-way power splitter. Eight of the outputs of the power splitter send the wideband noise signal to the eight ADC inputs of the CRS board through 36" cables, while the other eight outputs are connected to $50$-$\Omega$ terminations to minimize reflections from those signal paths.
}
\end{figure}

To validate the DSP performance of the CRS F-Engine implementation for CHORD, we present in this section benchtop measurements conducted with a single CRS board running the Pocket Correlator firmware. As described in Section~\ref{sec:scaler}, the sky signal can be modeled as a zero-mean Gaussian distribution. To represent such a noise-like signal for the purposes of validating the signal processing firmware, a wideband Gaussian noise source was generated using a $50$-$\Omega$ termination and analog passive and active components. The signal was amplified through two Mini-Circuits ZX60-43-S+ and two Mini-Circuits ZX60-30186-S+ amplifiers (along with a Mini-Circuits VAT-12A+ $12$~dB attenuator before the final amplifier to prevent compression) before being filtered with a Mini-Circuits VLF-1200+ low-pass filter, and split using a Mini-Circuits $16$-way ZC16PD-252-S+ power splitter, $8$~outputs of which are injected into the ADCs through a cable while the other $8$~outputs have $50$-$\Omega$ terminations to minimize reflections and improve isolation. This $16$-way splitter was selected for its 10--2500 MHz bandwidth, which covers the full first Nyquist zone (DC--$1600$~MHz) at the cost of higher loss due to additional splitting. We refer to each signal as ADC1--8, as indicated on the front panel of the CRS boards shown in Figure~\ref{fig:f_engine_crate}. A block diagram of the signal chain is shown in Figure~\ref{fig:signal_chain}.

\begin{figure} [h!]
\begin{center}
\begin{tabular}{c} 
\includegraphics[width = 0.9\linewidth]{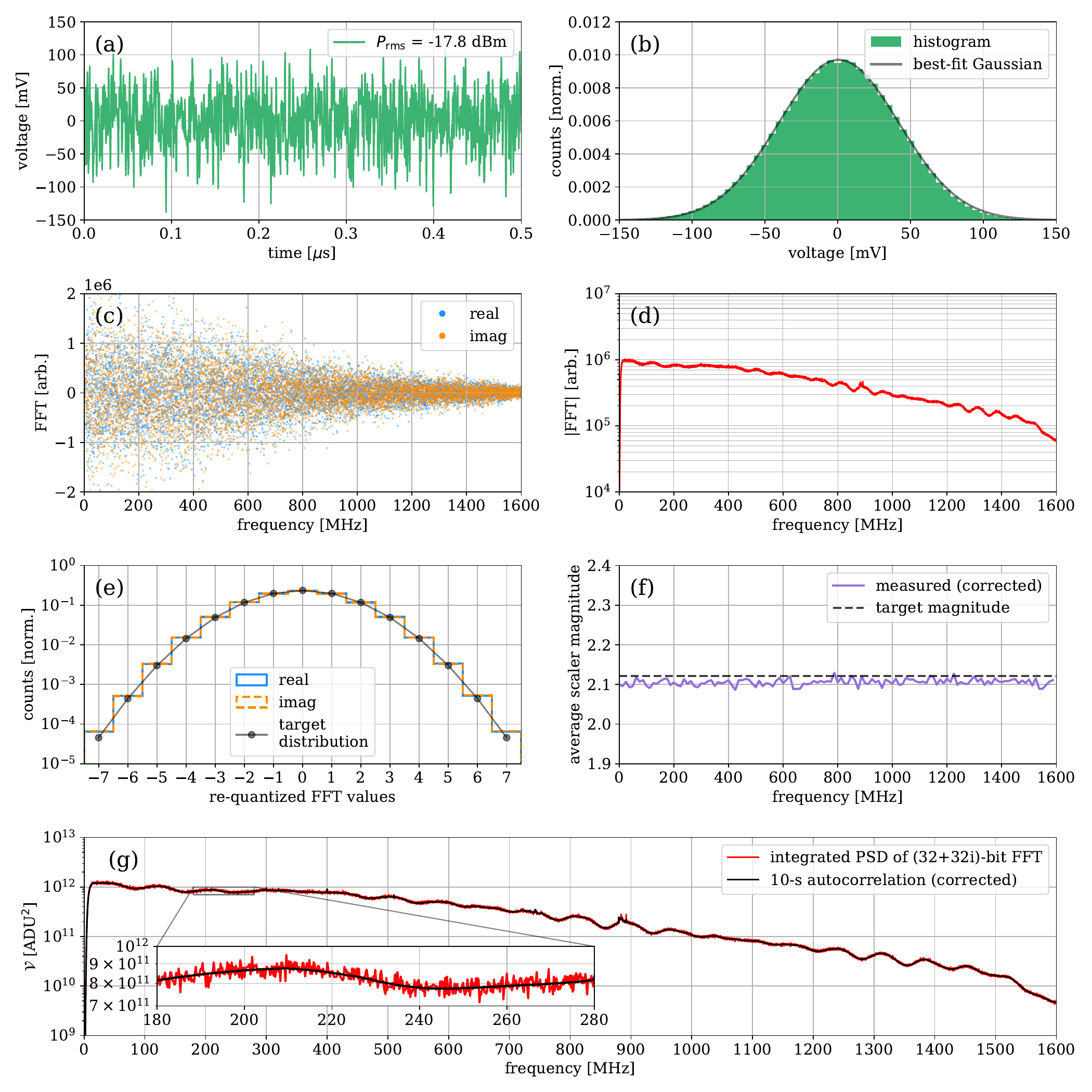}\end{tabular}
\end{center}
\caption[example] 
{ \label{fig:data_flow} 
Data products for key stages of the digital signal processing of \texttt{chFPGA}. All data shown in this figure correspond to ADC1. (a) The ADC timestream data for the first $0.5$~$\mu$s of a single frame. The RMS power $P_\text{rms}$ of the signal is indicated in the legend. (b) A histogram of the ADC timestream data, demonstrating its zero-mean Gaussian distribution. (c) The real (imaginary) part of the $(32 + 32i)$-bit FFT data in arbitrary units shown in blue (orange). (d) The average magnitude of $1000$~FFT frames. Panels~(c) and (d) show that our test setup produces power with a red spectrum, ripple from cable reflections, and minor RFI contamination (described further in the main text). (e) Histogram of the real (imaginary) parts of the $(4+4i)$-bit re-quantized FFT (``Scaler") data shown in blue (orange) as a function of the $15$ possible values from $-7$ to $+7$. The target distribution is shown as a black line with dots at each of these values for comparison. (f) The average magnitude of 1000 Scaler frames, corrected for quantization noise, is shown in purple, with the target magnitude shown as a dashed black line for comparison. (g) A single $10$-s autocorrelation product, also corrected for quantization noise, is shown as a solid black line, along with the power spectral density (PSD) as computed with $1000$~$(32+32i)$-bit FFT frames is shown in red for comparison. This figure demonstrates successful signal processing in agreement with expectations for each stage of the CRS F-Engine.}
\end{figure}

Figure~\ref{fig:data_flow} presents plots of the key data products described in Section~\ref{sec:firmware}. Given that all the signals receive a copy of the same wideband noise source, the measurements presented in Figure~\ref{fig:data_flow} are only shown for ADC1, whose results are representative of the other input channels.

The first row of Figure~\ref{fig:data_flow} presents measurements of the raw digitized ADC timestream of ADC1. Panel~(a) shows the voltage of the ADC signal in units of millivolts, converted from raw bits to voltage by noting that the full-scale of the ADC corresponds to $1$~V$_{\text{pp}}$. The ADC resolution is therefore $2^{\Delta} = 2^{14}$ levels, corresponding to $0.061$~mV/level. While the full length of a frame is $t_\text{frame} = 5.12~\mu$s, only the first $0.5~\mu$s are shown to avoid over-plotting points so that the noise structure of the signal can be more easily seen. Panel~(b) presents a histogram of ADC data from $1000$~frames, or a total of $5.12$~ms of data. The histogram is organized into $5$~mV bins for better visualization. The distribution of the signal levels is consistent with a zero-mean Gaussian, as expected for the amplified thermal noise produced by a $50$-$\Omega$ termination.

The second row of Figure~\ref{fig:data_flow} presents FFT measurements of the ADC data. Panel~(c) shows the real and imaginary components of a single $(32 + 32i)$-bit FFT frame across the first Nyquist zone for reference. Panel~(d) shows the average magnitude of $1000$ FFT frames, corresponding to a $5.12$~ms integration. The measurements display the expected structure of a wideband noise source amplified, filtered, and transmitted through a long cable, with the dominant features corresponding to (1) roll-off in the passband due to the amplifiers; (2) filtering; and (3) ripples caused by cable reflections. RFI begins to stand out in the spectra with increased integration, likely due to coupling with either one of the components of the signal chain or the CRS itself, the most prominent being the $869$--$894$~MHz Canadian cellular downlink band\cite{ised_srsp503_2024}. Further hints of other RFI sources or spurious tones begin to be apparent above the noise, but are not resolved more clearly until further integration is applied.

The third row of Figure~\ref{fig:data_flow} presents measurements of the re-quantized FFT after the digital gains have been computed and stored in the Scaler's BRAM as described in Section~\ref{sec:scaler}. Panel~(e) shows a histogram corresponding to the normalized number of counts for each re-quantized FFT value (from $-7$ to $+7$) over $1000$~Scaler frames, for both the real and imaginary parts of the data. The target distribution is shown for comparison as a semi-transparent black line with dots at each of the re-quantized FFT values. The target distribution is defined as a complex, circularly symmetric, zero-mean Gaussian signal whose width is set by the target magnitude used to compute the digital gains, and which does not account for saturated signals (i.e., the target distribution shown is a simplified model intended for comparison). The agreement between both the real and imaginary histograms and the target distribution demonstrates successful re-quantization of the FFT data. Further, the normalized counts corresponding to the saturation limits at -7 and 7 are suppressed by several orders of magnitude to acceptable levels (with the small difference between the measured values and the target distribution attributed to the simplified model used for the target distribution which does not account for saturations). 

Panel~(f) shows the average $(4 + 4i)$-bit Scaler magnitude over the same $1000$~frames as a function of frequency, with the median value computed across $128$~$64$-bin-wide chunks shown. The average Scaler magnitude is corrected for the inherent quantization noise introduced by the re-quantization of the FFT data. The overall consistency between the average Scaler magnitude and the target magnitude demonstrates that the digital gains are successfully applying the re-quantization of the $(32 + 32i)$-bit FFT data such that the quantization is acceptably optimized.

Finally, the bottom row of Figure~\ref{fig:data_flow}, corresponding to panel~(g), presents a roughly $10$-s autocorrelation of ADC1 using the built-in correlator firmware. {As with the average Scaler magnitude, this $10$-s autocorreltion is corrected for the inherent quantization noise introduced by the re-quantization of the FFT data.} The main panel shows that the overall structure of the autocorrelation traces the shape of the average FFT magnitude. The inset panel shows the same data zoomed-in on the $180$--$280$~MHz region to more effectively highlight the differences between the two lines. In particular, we see that the noise levels are significantly suppressed, as is expected due to the longer integration time. RFI corresponding to the $869$--$894$~MHz Canadian cellular downlink band is seen more clearly in comparison to the integrated FFT data, in addition to the fainter sub-bands of the $600$~and~$700$~MHz Canadian cellular bands\cite{ised_srsp518_2019}, along with several spurious tones. Overall, the results presented in this section demonstrate a successful initial validation of the DSP of the CRS F-Engine, with each data product in agreement with expectations.

\section{CONCLUSION}\label{sec:conclusion}

We introduced the CRS F-Engine, the first stage of the digital F-X correlator implementation of CHORD, a next-generation radio interferometer currently being deployed at the DRAO in Canada. The F-Engine is built using the t0.technology CRS, a multi-application microwave readout platform based on the AMD RFSoC which integrates a CPU, FPGA resources, and RF data converters all within the same integrated circuit. We first introduced the CRS hardware, which supports the required analog and digital signal processing at both the laboratory and full-deployment scales of $\mathcal{O}(100)$~CRS boards and over a thousand antenna inputs. Custom backplane-equipped crates enable distribution of a common clock and time synchronization signals and can implement a full-mesh $25$~Gbit/s network for intra-crate data transmission. We then presented the firmware capabilities of the \texttt{chFPGA} firmware platform, motivated by CHORD requirements. The CRS directly digitizes the $300$--$1500$~MHz CHORD bandwidth through interleaved ADC sampling while maintaining sub-sample phase synchronization across 128~CRS boards in the core array. It then efficiently channelizes the signals and re-quantizes the data to a lower bit-depth before transmitting the data over high-speed $100$~GbE fiber optic links to the X-Engine, which performs the spatial correlation computations. We finally demonstrated the capabilities of the CRS using the ``Pocket Correlator" firmware, which implements a fully FPGA-based F-X correlator for the 8~inputs of a single CRS. We validated the DSP performance of the CRS F-Engine using the Pocket Correlator firmware by injecting copies of a wideband Gaussian noise source into the CRS, and showed that all data products were consistent with expectations. 

The deployment of an initial array of $64$~CHORD dishes, the CHORD Pathfinder, is currently achieving rapid progress, expected to be complete within the third quarter of 2026. The Pathfinder F-Engine, corresponding to a single crate of $16$~CRS boards and its peripheral equipment, has been commissioned on-site alongside other CHORD systems. As more Pathfinder dishes come online, upcoming work will address the validation of the full end-to-end system as the F-Engine is further integrated with the CHORD analog and digital systems, in preparation for the full CHORD core array.

\acknowledgments       

I.H. is supported by Mitacs Accelerate Fellowship IT38715. M.D. is supported by a CRC Chair, NSERC Discovery Grant.

\bibliography{report} 
\bibliographystyle{spiebib} 

\end{document}